\documentclass[conference]{IEEEtran}
\usepackage[T1]{fontenc}
\usepackage{amsmath,amssymb}
\usepackage{algorithmic}
\usepackage{graphicx}
\usepackage{booktabs}
\usepackage{array}
\usepackage{placeins}
\usepackage{url}
\usepackage{tikz}
\usepackage{float}
\usepackage[numbers,sort&compress]{natbib}
\usetikzlibrary{arrows.meta,positioning}
\usepackage[hidelinks]{hyperref}

\newcommand\ourdataset{{\textrm{HandSchA}}}

\newcommand{\G}{\mathcal{G}}
\newcommand{\V}{\mathcal{V}}
\newcommand{\E}{\mathcal{E}}

\begin{document}
\title{A Novel Dataset and Framework for Schenkerian Optical Music Recognition}

\author{
Yixuan Huang\IEEEauthorrefmark{1},
Xiao Yun\IEEEauthorrefmark{2},
Jiayin Zheng\IEEEauthorrefmark{3},\\
Simon Mak\IEEEauthorrefmark{1},
Yue Jiang\IEEEauthorrefmark{1},
Cynthia Rudin\IEEEauthorrefmark{1},
\IEEEauthorblockN{Stephen Ni-Hahn\IEEEauthorrefmark{3}}
\IEEEauthorblockA{\IEEEauthorrefmark{1}Duke University
\IEEEauthorrefmark{2}Independent Researcher, Denton, USA
\IEEEauthorrefmark{3}Chinese University of Hong Kong, Shenzhen\\
\texttt{yixuan.huang@duke.edu}, \texttt{stephennihahn@cuhk.edu.cn}}}
\maketitle
\begin{abstract}
Music theory enables performers, composers, and listeners to engage more deeply with music. Schenkerian analysis (SchA) is a particularly illustrative and widely used approach for modeling tonal structure, organizing music into hierarchical layers of harmonic and voice-leading relationships. Despite its importance in pedagogy and scholarship, producing well-formatted Schenkerian graphs remains difficult. 
A typical workflow might involve writing the Schenkerian analysis on paper, and then manually, painstakingly, entering each symbol by hand into music software, exporting 
to vector graphics, and manually refining layouts in graphical editors. Existing optical music recognition (OMR) systems for handwritten \textit{common western music notation} (CWMN) can be leveraged for their ability to recognize small-scale symbols common to those in SchA. However, CWMN OMR models generally fail to recognize large symbols such as phrase slurs, struggle with dense notation, and do not recognize Schenkerian analysis markings. Thus, these models do not extend well to digitizing SchA. This paper introduces a novel framework for optical music recognition of both CWMN and SchA. Our novel contributions include (1) an ensemble model in which each individual model follows a different image-cropping scheme for improved object detection, (2) a topological neural network architecture for notation assembly, where candidate hyperedges attend to one another and perform hierarchical message passing, (3) the first dataset of labeled handwritten SchAs, gathered with novel model integration to the OMR-annotation software, MuNG Studio, and (4) the first algorithm for converting semantic Schenkerian graphs into typeset format.

\end{abstract}
\begin{IEEEkeywords}
Optical music recognition, topological neural networks, Schenkerian analysis, handwritten music datasets
\par\textit{Repository}---{\mdseries\href{https://github.com/stephenHahn88/mung-studio-schenker}{github.com/stephenHahn88/mung-studio-schenker}}
\end{IEEEkeywords}

\section{Introduction}
\label{sec:introduction}

Music theory provides frameworks to understand the structure and form of musical works, enabling performers, composers, and listeners to develop a deeper connection with music. In particular, Schenkerian analysis (SchA) offers a powerful and widely studied approach for modeling tonal structure, organizing music into hierarchical layers of harmonic and voice-leading relationships \citep{schenker2001free}. Such analyses are fundamental to pedagogy, performance, and composition, offering insight into how harmonic patterns unfold over time. 

Unfortunately, typeset Schenkerian analyses remain prohibitively difficult to produce, share, and adapt; typesetting involves a multi-stage workflow with significant manual effort and specialized knowledge. One generally starts on paper, and then encodes as much information as possible in music notation software (e.g., Dorico or Sibelius \citep{dorico,AvidSibelius}), exports the result to a vector format such as SVG, and then manually refines the graphical layout using tools such as Adobe Illustrator. Often, Schenkerian graphs are then shared as static images (e.g., JPG or PNG). As a result, access to well-formatted, editable Schenkerian analyses remains heavily limited. Furthermore, any computer-readable semantics recorded in music notation software (e.g., symbol classes and pitch information) are lost once music is exported to SVG. Given promising work on leveraging Schenkerian structure for composition, analysis and performance \cite{marsden2010schenkerian,kirlin2015learning,dodson2008performance,ni2026autoscha,ni2026progress,hahn2023melody,hahn2023neworleans}, an automated workflow for Schenkerian typesetting is essential for automating such tasks. 

In parallel, optical music recognition (OMR) for common western music notation (CWMN) has advanced considerably, enabling the transcription of handwritten notation into typeset formats. However, existing OMR systems still struggle with large symbols and dense notation, and are not suited for Schenkerian graphs. Schenkerian analysis extends CWMN with more complex and ``stretched'' notation, including widely varying stem lengths, nested slurs that can span entire pages, and various levels of harmonic analysis. These elements interact in ways that are both visually complex and semantically dependent on broader musical context, rendering traditional OMR approaches, which rely only on local-window symbol detection and edge relationships, insufficient.

\begin{figure}
    \centering
    \includegraphics[width=\linewidth]{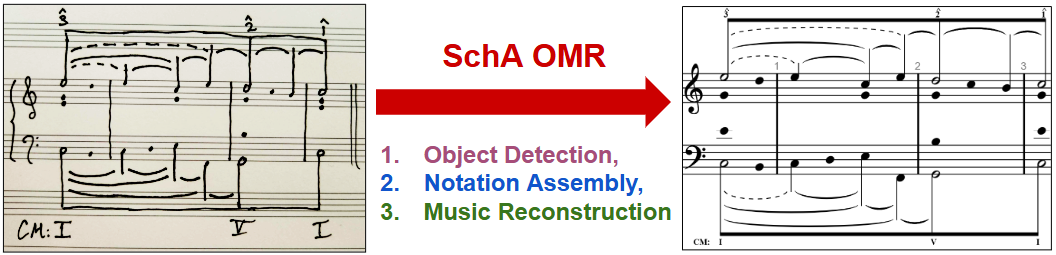}
    \caption{Handwritten SchA (left) transformed into a typeset graph (right).}
    \label{fig:omr_example}
\end{figure}

Recognizing individual symbols is only the first step toward recovering a machine-readable Schenkerian analysis. The detected symbols must also be assembled into a semantic graph that represents their musical relationships and hierarchical organization. This reconstruction is challenging because Schenkerian notation combines small, densely arranged symbols with extended markings whose interpretation depends on relationships spanning substantial portions of the page. Consequently, recognition must account for both variation in visual scale and dependencies among candidate relationships.

We address these challenges through a recognition-to-typesetting framework for handwritten SchA (Figure~\ref{fig:omr_example}), whose symbol recognition and syntax-link prediction also apply to CWMN. The framework combines complementary symbol detectors with a notation-assembly stage in which a topological neural network predicts syntax links and gradient-boosting classifiers predict time groups and precedence links. In this network, candidate hyperedges interact through attention to determine context representations for each symbol. Final link predictions are determined through message passing between node, edge, and context embeddings. The resulting semantic representation supports state-of-the-art performance on automated typesetting. Our novel contributions include:

\begin{enumerate}
    \item A complementary YOLO-DETR object-detection ensemble using different image-cropping schemes to address variation in symbol scale,

    \item A topological neural network for notation assembly, combining attention among candidate hyperedges with hierarchical message passing,

    \item A labeled dataset of handwritten Schenkerian analyses, developed using model-assisted annotation features integrated into MuNG Studio~\citep{mungstudio},

    \item An algorithm for converting semantic Schenkerian graphs into
    typeset notation.
\end{enumerate}

Section \ref{sec:schenkerian_notation} introduces the necessary background on SchA. Section \ref{sec:prelims_related_work} discusses prior work on OMR datasets, labeling tools, and models. Section \ref{sec:dataset} describes and analyzes our novel \ourdataset{} dataset. Section~\ref{sec:data-collection} describes our data collection and model update scheme. Section~\ref{sec:method} presents symbol recognition, syntax-link prediction, and semantic conversion. Section~\ref{sec:experiments} evaluates symbol recognition and syntax-link prediction on Schenkerian and standard music notation.

\section{Preliminaries on Schenkerian Notation}
\label{sec:schenkerian_notation}

\begin{figure}[t]
    \centering
    \includegraphics[width=\linewidth]{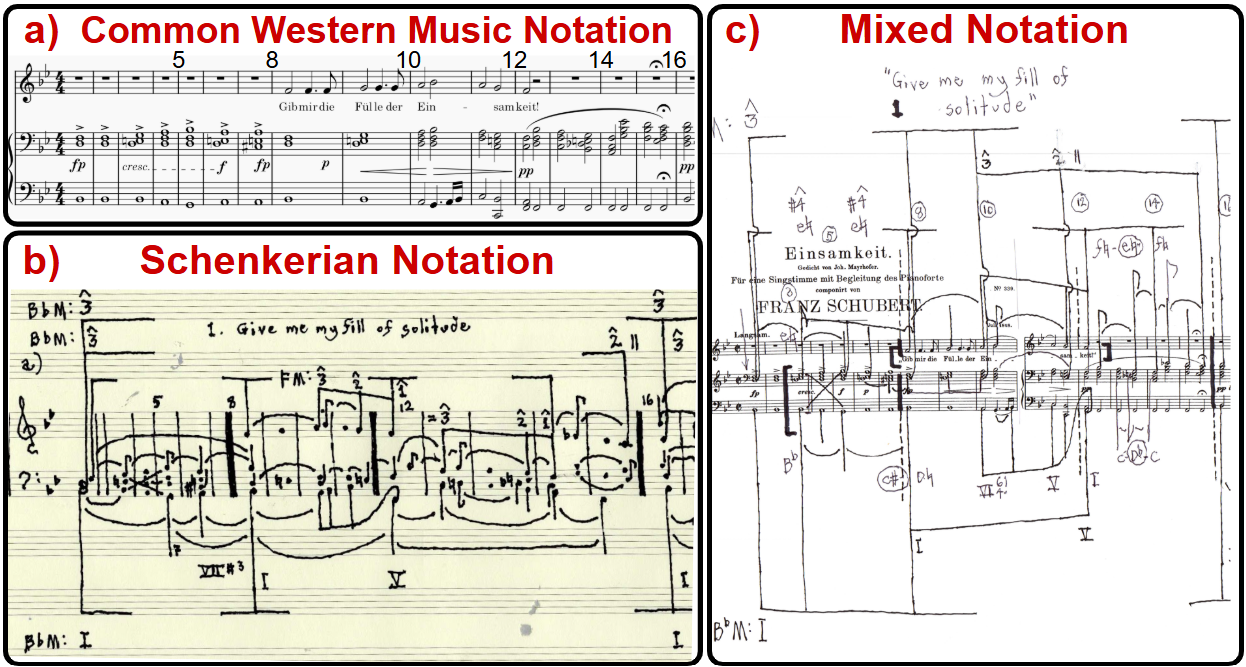}
    \caption{a) Typeset CWMN, b) handwritten SchA, and c) mixed notation, from Timothy Jackson's analysis of Schubert's \textit{Einsamkeit}, D. 620, mm. 1-16.}
    \label{fig:notation_example}
    \vspace{-5mm}
\end{figure}

Schenkerian notation is specialized for representing hierarchical tonal structure, rather than precise performance instructions; it abstracts away many details to emphasize underlying voice-leading relationships and harmonic structure. There are two common ways of displaying SchA: 1) ``pure'' notation (Figure \ref{fig:notation_example}b), which abstracts away pitches in favor of structural clarity; and 2) mixed notation (Figure \ref{fig:notation_example}c), which overlays structural annotations directly on a CWMN score. Often, an analyst will do initial work directly on the score, then transfer the most important aspects to a pure SchA graph. We will focus on pure notation, but ensure extendability to mixed notation. For the remainder of the paper, SchA notation refers to pure graphs unless otherwise specified.

Visually, SchA employs symbols similar to CWMN, but these elements often serve different semantic roles (e.g., noteheads, stems, and beams representing levels of hierarchy rather than rhythmic duration, stem length drawing attention to structural tones, or slurs describing long-range connections rather than phrasing or articulation). There also exist unique SchA notation elements (e.g., key and Roman numeral/figured bass analysis being associated with various levels of structure, hats over numbers representing hierarchies of scale degrees associated with structural harmonies, beams sometimes being stretched and curved to indicate a harmony \textit{unfolding} over time, or structures being marked by an \textit{interruption} (e.g., ``||'' in the upper voice after m. 15), indicating a halting of the music's trajectory at a particular hierarchical level.

From an OMR perspective, these differences introduce several challenges. First, the symbol vocabulary only partially overlaps with CWMN, requiring redefinition of classes and labels. Second, the meaning of relationships between symbols is fundamentally different, emphasizing hierarchy and dependency rather than instructions for performance. Finally, the layout is less standardized, with greater variability in spacing and alignment. These characteristics make Schenkerian notation a distinct and underexplored domain for OMR, necessitating specialized datasets, annotation schemes, and modeling approaches.

\section{Related Work}
\label{sec:prelims_related_work}

The development of optical music recognition (OMR) systems has historically been constrained by the availability of high-quality datasets. Among handwritten music datasets, MUSCIMA++ \citep{muscimadataset} remains one of the most widely used benchmarks, providing symbol-level annotations and explicitly encoding relationships between musical primitives through a graph-based representation. The dataset is derived from the CVC-MUSCIMA corpus~\cite{fornes2012cvcmuscima} and includes multiple writers, supporting variability in handwriting. DoReMi~\cite{shatri2021doremi} provides printed scores generated with notation software, together with symbol and relation annotations. More recent work has sought to address limitations in scale and diversity through synthetic data generation. Smashcima \citep{smashcima} introduces a full-page handwritten music document synthesizer capable of producing structurally consistent scores with controllable variation. Such synthetic approaches are particularly valuable for training data-hungry models, though they may lack the complexity and irregularity of real manuscripts. Despite these advances, existing datasets are largely focused on CWMN. They do not capture alternative analytical notations, such as SchA, motivating the need for new datasets tailored to such specialized domains.

The Music Notation Graph (MuNG) formalism has emerged as a key representation for structured annotation, particularly through its use in the MUSCIMA++ dataset \citep{muscimadataset}. MuNG models notation as a graph in which low-level primitives (e.g., noteheads, stems) are nodes and musical relationships are explicit edges, thereby bridging symbol detection and higher-level musical semantics. This enables richer supervision than flat label or bounding-box annotations, but also substantially increases annotation complexity. Early tools such as MUSCIMarker~\cite{hajic2017muscimarker} supported quick creation of MuNG annotations, yet suffered from limited usability and maintainability. MuNG Studio addressed this gap by providing a web-based environment for viewing and editing MuNG annotations \citep{mungstudio}. Our work extends the capabilities of MuNG Studio to include SchA node classes, described further in Section \ref{sec:dataset}.

OMR systems are often divided into two categories: end-to-end and multi-stage. Recent end-to-end models typically treat OMR as an image-to-sequence problem, directly producing symbolic encodings (e.g., MusicXML or Humdrum **kern) from staff, system, or page images. These approaches work with a variety of settings, from monophonic and camera-like recognition \citep{calvo2018camera} to polyphonic staff-level recognition \citep{edirisooriya2021empirical}. Transformer-based systems extend this image-to-sequence direction. TrOMR \citep{li2023tromr} applies a Transformer architecture to polyphonic OMR and real-world camera scenarios, Sheet Music Transformer \citep{rios2024sheet} extends transcription beyond monophonic settings, and pianoform OMR work investigates MusicXML-compatible encodings and full-page transcription \citep{mayer2024practical,rios2025fullpage}. 

These systems have achieved strong results on increasingly complex CWMN settings, but they generally require large paired image and symbolic datasets built around sequential representations. For example, Camera-PrIMuS contains 87,678 paired image and sequence incipits \citep{calvo2018camera}; TrOMR reports about 0.4 million rendered staff images and 71,292 camera-derived staff images \citep{li2023tromr}; and Sheet Music Transformer uses GrandStaff~\cite{rios2023endtoend} with 53,882 pianoform systems and a Quartets corpus of 38,051 excerpts \citep{rios2024sheet}. %

In parallel, multi-stage OMR systems have proven highly capable, typically dividing the full process into four stages: preprocessing, music object detection, notation assembly, and music encoding \citep{rebelo2012optical,calvo2020understanding}. This object-level, structurally aware direction is particularly useful when labeled data are scarce, because preliminary detector outputs can assist human annotation before a fully automatic end-to-end system is available (see Section~\ref{sec:dataset}). Preprocessing typically involves binarization and staff removal. For music object detection, Pacha et al.~\cite{pacha2018baseline} establish an early deep-learning baseline by comparing Faster R-CNN, RetinaNet, and U-Net variants across multiple datasets. Shatri and Fazekas~\cite{shatri2024knowledge} compare Faster R-CNN detection with Mask R-CNN instance segmentation on full pages of DoReMi. In a more complete pipeline, Yang et al.~\cite{yang2024complete} use a YOLOv8-based detector on MUSCIMA++ and train notation assembly directly on detector outputs. Ultralytics YOLOv8 \citep{yolov8_ultralytics} provides a general-purpose detector implementation. Luo et al.~\cite{luo2024mdetr} propose M-DETR, a Transformer-based detector adapted for OMR. Penarrubia et al.~\cite{penarrubia2026musvit} pretrain MuSViT, a vision Transformer for sheet music, and evaluate symbol detection on DeepScoresV2~\cite{tuggener2018deepscores,tuggener2021deepscoresv2} with a Faster R-CNN head. Tuggener et al.~\cite{tuggener2024realworld} adapt a detector trained on DeepScoresV2 to real scans of printed scores with augmentation and unsupervised domain adaptation~\cite{castellanos2021unsupervised}.

For notation assembly and structural transcription, Pacha et al.~\cite{pacha2019learning} formulate MuNG assembly as a learned graph construction problem over detected primitives, and MusiGraph by Bar{\'o} et al.~\cite{baro2022musigraph} combines object detection with graph neural networks. MUSCIMA++ annotates syntactic relationships only and leaves precedence and simultaneity to be inferred \citep{muscimadataset}. The OmniOMR annotation guidelines~\cite{omniomr_annotation_guidelines} add precedence links as a separate, manually annotated layer, and existing learned assembly methods predict syntax links only. Efficient Notation Assembly~\cite{penarrubia2023efficient} and completeOMR~\cite{yang2024complete} score each candidate pair from symbol classes and box geometry. Like the method of Pacha et al.~\cite{pacha2019learning}, these two methods judge each candidate pair on its own, even though candidates at the same symbol often constrain one another. A notehead, for example, usually attaches to a single stem. GNNs can pass context between symbols. Models such as GraphSAGE~\cite{hamilton2017graphsage} first aggregate information from nearby symbols and then score a candidate from the representations of its two endpoints. The candidate links themselves have no representations in such models, so they cannot exchange information directly.

Topological deep learning treats relations as objects with representations of their own. In a combinatorial complex, cells of different ranks pass messages to each other along boundary and incidence relations~\cite{hajij2022topological}. TopoTune~\cite{papillon2025topotune} further shows that once a complex is expanded into several graphs according to its neighborhood relations, any graph neural network can serve as the message-passing module. We adopt this view and treat symbols as 0-cells and candidate links as 1-cells. Messages travel only along the incidences between them, and attention serves as a message-passing module. Candidate links that share a symbol can therefore exchange information through that symbol.

\section{The \ourdataset{} Dataset}
\label{sec:dataset}

To support the training and evaluation of Schenkerian OMR models, we introduce \ourdataset{}, the first dataset of labeled handwritten SchAs. It contains 47 pages of handwritten Schenkerian graphs by 3 expert analysts covering 12 works by Monteverdi, Pachelbel, Bach, Schubert, Rachmaninov, and Berg, with 30,780 annotated symbols in 128 classes. 

\subsection{Annotation Scheme}
\label{sec:annotation-scheme}

Each page is annotated as a MuNG graph~\cite{muscimadataset}, in which every symbol is a node with a class label and a bounding box. Unlike the original MuNG vocabulary, we make finer distinctions for some symbols, such as the upward and downward curvature of slurs, because the curvature of a slur indicates voice relationships in SchA.

For most links, we follow the syntax and precedence conventions of the OmniOMR annotation guidelines~\cite{omniomr_annotation_guidelines}. Roman numeral analysis and linear progressions are handled similarly to lyrics in standard MuNG. Beyond these conventions, to record the voice and structural level of each analytical symbol, we also link structural stems to the analytical symbols they mark, including slurs, beams, the Roman numerals of the harmonic analysis, and the numerals of linear progressions. Beams are also linked to the harmonic-analysis and linear-progression labels they span. Further details are documented online in the GitHub repository \url{https://github.com/stephenHahn88/mung-studio-schenker}.

\subsection{Data Collection}
\label{sec:data-collection}

\begin{figure}[t]
    \centering
    \includegraphics[width=0.75\linewidth]{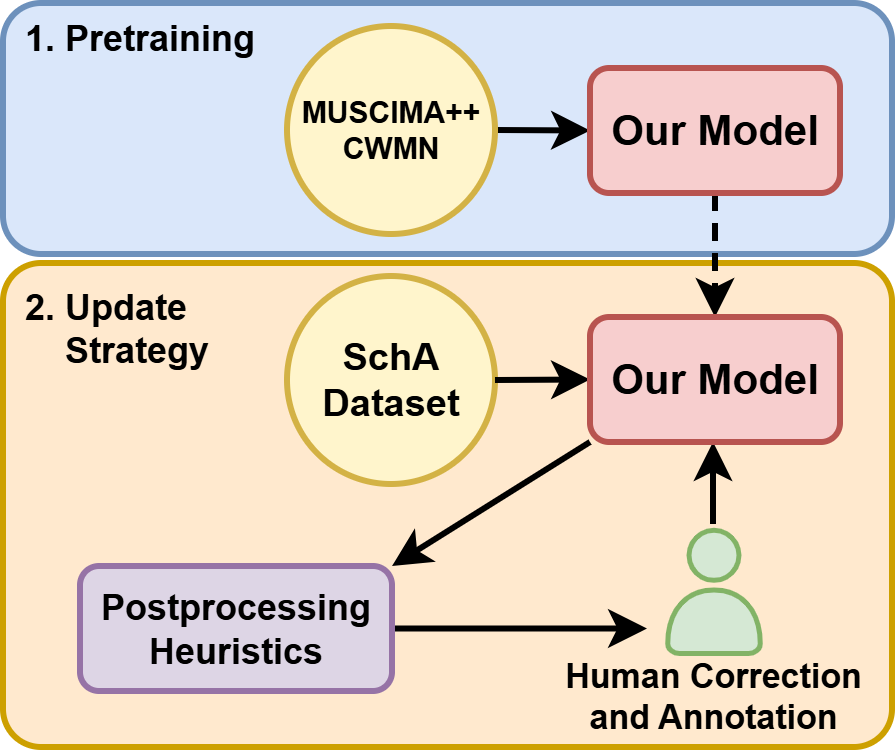}
    \caption{Model pretraining and update strategy for data collection. A symbol detector is first pretrained on CWMN data (top). The symbol detector and the prediction models for syntax links, time groups, and precedence links then generate initial predictions for new SchA pages. Predictions are postprocessed with heuristics and then corrected and completed by a human annotator. The corrected pages become new training data for these models (bottom).}
    \label{fig:data_steps}
\end{figure}

We collect annotations with a model-assisted, multi-stage process (Figure~\ref{fig:data_steps}). At first, we did not have enough Schenkerian data to train a detector that performs well, so we pretrained a symbol detector on MUSCIMA++~\cite{muscimadataset}, a dataset of handwritten CWMN scores. Because SchA extends CWMN, the pretrained detector already gave reasonable bounding boxes for most symbols on Schenkerian pages. We mapped CWMN classes to their Schenkerian counterparts and postprocessed the predictions with heuristics so that they were easier to manually correct. Starting from this detector, we fine-tuned it on a small number of annotated Schenkerian pages and used the fine-tuned detector to generate initial symbol predictions for new pages. As more pages were annotated, we also trained separate prediction models for syntax links, time groups, and precedence links (Section~\ref{sec:method}), which give initial predictions of these three relations on new pages. 

Annotators work in MuNG Studio~\cite{mungstudio}, which we extend with Schenkerian node classes and integrate with our models. An annotator can run the models on any region of a page and then correct and complete their predictions by fixing classes and bounding boxes, adding missing symbols, and adding or revising syntax links, time groups, and precedence links. Each corrected page is added to the training set, and the four retrained models produce the initial predictions for subsequent pages, completing the update loop in Figure~\ref{fig:data_steps}.

\subsection{Dataset Statistics and Splits}
\label{sec:dataset-splits}

\begin{table}[!htb]
\centering
\caption{Subsets of \ourdataset{} used in our experiments. Annotations are symbols in the symbol-recognition subset and the named relation in the other subsets.}
\label{tab:dataset-split}
\newcommand{\splithead}[2]{\begin{tabular}[b]{@{}c@{}}#1\\#2\end{tabular}}
\begin{tabular*}{\linewidth}{@{\extracolsep{\fill}}lrrrr@{}}
\toprule
 & \splithead{Symbol}{recognition} & \splithead{Syntax}{links} & \splithead{Time}{groups} & \splithead{Precedence}{links}\\
\midrule
Pages & 47 & 36 & 37 & 44\\
Annotations & 30,780 & 22,138 & 2,309 & 11,038\\
\bottomrule
\end{tabular*}
\vspace{-3mm}
\end{table}

\begin{table}[!htb]
\centering
\caption{Training, validation, and test pages in each split.}
\label{tab:dataset-partitions}
\newcommand{\splithead}[2]{\begin{tabular}[b]{@{}c@{}}#1\\#2\end{tabular}}
\begin{tabular*}{\linewidth}{@{\extracolsep{\fill}}lcccc@{}}
\toprule
Split & \splithead{Symbol}{recognition} & \splithead{Syntax}{links} & \splithead{Time}{groups} & \splithead{Precedence}{links}\\
\midrule
Random splits & 30/7/10 & 23/5/8 & 24/5/8 & 28/7/9\\
Held-out & 33/7/7 & 26/5/5 & 25/5/7 & 30/7/7\\
\bottomrule
\end{tabular*}
\vspace{-5mm}
\end{table}

Table~\ref{tab:dataset-split} gives the size of the data used in our experiments. For each subset, we report the average over eight random splits. We also report the performance on a separate split that holds out one work. Random splits assign pages to training, validation, and test at random. The held-out split tests on all pages of Rachmaninov's \textit{The Bells}, draws the same number of validation pages as the random splits at random from the remaining pages, and trains on the rest. Table~\ref{tab:dataset-partitions} gives the page counts of each split.

\section{Methodology}
\label{sec:method}

Our goal is to convert a page of handwritten music notation to typeset notation. To accomplish this task, we must first represent a page of handwritten SchA as a semantic graph $\G=(\V,\E_s,\E_p,\mathcal{H})$ through symbol recognition and notation assembly. We choose this multi-stage design over an end-to-end model so that the models can already assist annotators when only a few pages are labeled (Section~\ref{sec:data-collection}). Each node in $\V$ corresponds to a visible musical symbol. Syntax links $\E_s$ connect related symbol components such as noteheads and stems, while precedence links $\E_p$ specify the musical order of notes and analytical symbols. Each time group in $\mathcal{H}$ is a hypernode that contains noteheads that occur simultaneously in the reduction. We then convert a semantic graph into standardized JSON notation for Schenkerian typesetting \cite{nihahn2024newdatasetnotationsoftware} (see Algorithm~1).

\subsection{Symbol Recognition}
\label{sec:symbol-recognition}

\begin{figure*}[t]
  \centering
  \includegraphics[width=\textwidth]{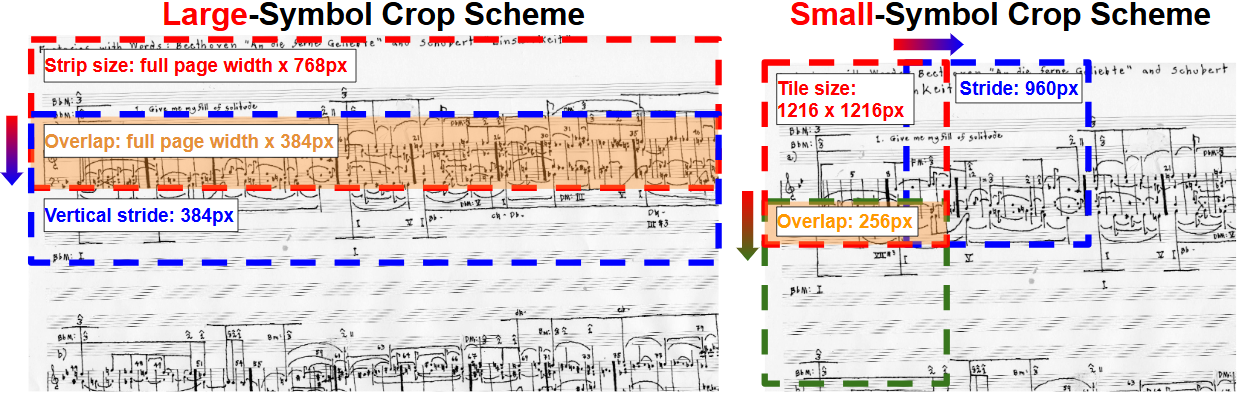}
  \caption{Crop layouts for symbol recognition: RF-DETR detects large symbols
  in overlapping full-width horizontal strips (left), while YOLOv8L detects small
  symbols in overlapping square tiles (right).}
  \label{fig:symbol-detection}
  \vspace{-5mm}
\end{figure*}

Written music often contains objects at vastly different scales, from small dots altering note durations to phrase slurs spanning entire pages. Schenkerian notation takes this dichotomy to the extreme. Small symbols such as noteheads, accidentals, and scale-degree marks require high local resolution, while long slurs, structural beams, and extended stems require larger context. We propose a novel ensemble model and cropping scheme combining YOLOv8L~\cite{yolov8_ultralytics} on overlapping square tiles with RF-DETR~\cite{robinson2026rfdetr} on overlapping full-width horizontal strips (Figure~\ref{fig:symbol-detection}) to capture both large- and small-scale symbols.

Following the image patch strategy of completeOMR \cite{yang2024complete}, we detect smaller local symbols using YOLOv8L. 
Neighboring tiles overlap, so a symbol near a tile boundary appears in more than one tile. We divide the page into non-overlapping regions, one per tile, with each boundary halfway between two neighboring tile centers. At inference time, a tile keeps a prediction only if the center of the predicted box lies in its region. A symbol in an overlap therefore normally comes from a single tile.

The large-symbol branch divides the page into overlapping full-width horizontal strips. We measure the size of a symbol by $L$, the longest side of its bounding box in pixels after the page is scaled to $5100\times3509$, the most common page size in \ourdataset{}. RF-DETR learns from large-symbol annotations in these strips, retaining horizontal context for slurs and beams that span several local tiles. Vertical overlap reduces truncation at strip boundaries. Both branches use their respective crop layouts during training and inference and output a symbol class, bounding box, and confidence score between 0 and 1 for each detection.

We merge the predictions of the two branches in three steps. First, we keep RF-DETR boxes with confidence at least 0.2 and apply class-wise non-maximum suppression at an intersection over union (IoU) of 0.5 within each branch. Second, we remove fragments of long symbols cut by tile edges. A YOLOv8L box within 128 pixels of an interior tile edge is dropped when at least 80\% of it lies inside an RF-DETR box of the same class, that RF-DETR box has $L>400$, and the YOLOv8L box has at most 80\% of its area. Third, we apply the same NMS to the boxes of both branches together. The resulting classes and page-level boxes form the symbol nodes $\V$, which provide the input to notation assembly.

\subsection{Syntax Links}
\label{sec:syntax-predictor}
Syntax links $\E_s$ describe how notation components relate to one another. Given the symbol classes and bounding boxes that symbol recognition produces for a page, we first estimate whether each candidate pair is connected and then assign its direction. To estimate connections, we use a topological neural network that combines the classes and boxes of the two symbols in each candidate pair with messages exchanged among candidate links that share a symbol.

\begin{figure}
\centering
\includegraphics[width=0.44\textwidth]{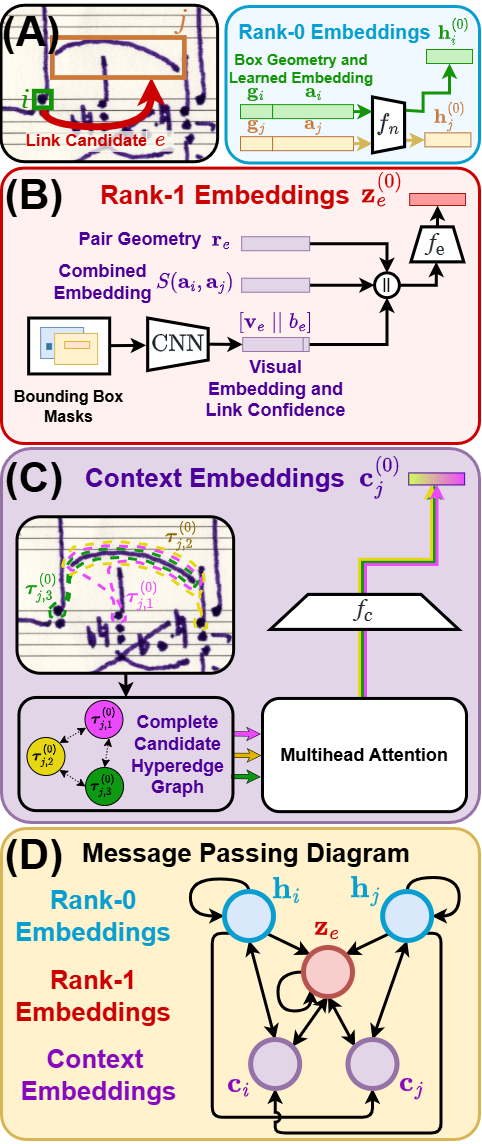}
\caption{Syntax-link prediction through symbol--candidate incidences. \textbf{(A-left)} Rank-0 cells (0-cells) $i$ and $j$ represent a notehead and slur respectively. The rank-1 cell (1-cell) connects $i$ to $j$. \textbf{(A-right)} The first rank-0 embeddings $\mathbf{h}^{(0)}$ are determined via box geometry $\mathbf{g}$ and learned class embeddings $\mathbf{a}$, fed through a shared encoder $f_n$. \textbf{(B)} The binary masks of the two boxes provide visual features $\mathbf v_e$ and a confidence logit $b_e$. A shared encoder processes the visual features, combined class embeddings $S(\mathbf a_i,\mathbf a_j)$, and pair geometry $\mathbf r_e$ to create initial rank-1 embeddings $\mathbf{z}_e^{(0)}$. \textbf{(C)} Initial context embeddings $\mathbf c_j^{(0)}$ associated with each object are generated through multihead attention with $K$ channels over all candidate hyperedges $\boldsymbol{\tau}_{j,e}$ associated with node $j$ and edge $e$. \textbf{(D)} Tensor diagram summarizing the message passing scheme between node, edge, and context embeddings.}
\label{fig:syntax-encoding}
\end{figure}

\begin{table}[!t]
\centering
\caption{Geometry used by the syntax-link model. Pair geometry is unchanged when the endpoints are exchanged. Incidence geometry describes one endpoint relative to the other.}
\label{tab:syntax-geometry}
\footnotesize
\renewcommand{\arraystretch}{1.16}
\setlength{\tabcolsep}{3pt}
\begin{tabular}{@{}llc@{}}
\toprule
Feature group & Definition & Dim. \\
\midrule
\multicolumn{3}{@{}l}{\textbf{Box
geometry $\mathbf g_i$}} \\
Box corners & $\ell_i/W,\ t_i/H,\ r_i/W,\ b_i/H$ & 4 \\
Center & $x_i/W,\ y_i/H$ & 2 \\
Size & $w_i/W,\ h_i/H$ & 2 \\
\midrule
\multicolumn{3}{@{}l}{\textbf{Symmetric pair geometry $\mathbf r_e$}} \\
Center offsets & $|\Delta x|/W,\ |\Delta y|/H$ & 2 \\
Center distance & $\sqrt{\Delta x^2+\Delta y^2}/D$ & 1 \\
Box distance & $\sqrt{d_x^2+d_y^2}/D$ & 1 \\
Projection gaps & $d_x/W,\ d_y/H$ & 2 \\
Projection overlaps & $o_x/\min(w_i,w_j),\ o_y/\min(h_i,h_j)$ & 2 \\
Enclosing box & $u_w/W,\ u_h/H$ & 2 \\
Size ratios & $\min(w_i,w_j)/\max(w_i,w_j),$ & \\
 & $\min(h_i,h_j)/\max(h_i,h_j)$ & 2 \\
Absolute log size ratios & $|\log(w_i/w_j)|,\ |\log(h_i/h_j)|$ & 2 \\
Intersection over union & $o_xo_y/(A_i+A_j-o_xo_y)$ & 1 \\
Log area ratio & $|\log(A_i/A_j)|$ & 1 \\
\midrule
\multicolumn{3}{@{}l}{\textbf{Incidence geometry $\boldsymbol\delta_{i\to j}$}} \\
Page-scaled offsets & $\Delta x/W,\ \Delta y/H$ & 2 \\
Symbol-scaled offsets & $\Delta x/\sqrt{w_iw_j},\ \Delta y/\sqrt{h_ih_j}$ & 2 \\
Signed log size ratios & $\log(w_j/w_i),\ \log(h_j/h_i)$ & 2 \\
Horizontal edge offsets & $(\ell_j-r_i)/W,\ (r_j-\ell_i)/W$ & 2 \\
Vertical edge offsets & $(t_j-b_i)/H,\ (b_j-t_i)/H$ & 2 \\
\bottomrule
\end{tabular}
\par\vspace{4pt}
\begin{minipage}{\linewidth}
\footnotesize
The page has width $W$, height $H$, and diagonal $D=\sqrt{W^2+H^2}$. Box $i$ has corners $(\ell_i,t_i)$ and $(r_i,b_i)$, center $(x_i,y_i)$, size $(w_i,h_i)$, and area $A_i=w_ih_i$. We use $\Delta x=x_j-x_i$ and $\Delta y=y_j-y_i$. The projection gaps $d_x,d_y$ are zero for touching or overlapping intervals, and $o_x,o_y$ are their overlap lengths. The smallest box enclosing both endpoints has size $(u_w,u_h)$.
\end{minipage}
\vspace{-5mm}
\end{table}

\vspace{2mm}
\noindent\textbf{Finding candidate pairs.}
For each symbol, we select its 64 nearest neighbors by the minimum Euclidean distance between bounding boxes. Boxes that overlap or touch have distance zero. If either $i$ selects $j$ or $j$ selects $i$ as a candidate, unordered pair $e = \{i,j\}$ enters the candidate set. We write $\mathcal I_i$ for the candidates that contain symbol $i$.

\vspace{2mm}
\noindent\textbf{Visual and geometric encoding.}
For each candidate pair, we draw a binary mask for each of the two symbol boxes and use the two masks as input channels. Following the link prediction method of Pacha et al.~\cite{pacha2019learning}, a CNN processes these two channels with five convolutional blocks and global average pooling, producing a visual embedding vector $\mathbf v_e$ and a connection confidence logit $b_e$. We average the features obtained with the two mask orders so that swapping endpoints preserves the representation.

Besides the CNN outputs, the model encodes symbol classes and geometry. Each symbol class has a learned embedding vector, and $\mathbf a_i$ is the embedding of the class of symbol $i$. Table~\ref{tab:syntax-geometry} lists all geometric features, where $\mathbf g_i$ describes the box of symbol $i$ and $\mathbf r_e$ describes the relation between the two boxes.
We provide a combined representation of two embeddings as $S(\boldsymbol{\zeta},\boldsymbol{\xi})=\bigl[\boldsymbol{\zeta}+\boldsymbol{\xi} \Vert |\boldsymbol{\zeta}-\boldsymbol{\xi}|\bigr]$, which concatenates their sum and absolute difference. Here $\Vert$ denotes concatenation. The initial symbol and candidate states are
\begin{align*}
 \mathbf h_i^{(0)}&=f_n\Bigl(\bigl[\mathbf a_i\Vert\mathbf g_i\bigr]\Bigr),\\
 \mathbf z_e^{(0)}&=f_e\Bigl(\bigl[\mathbf v_e\Vert b_e\Vert\mathbf r_e\Vert S(\mathbf a_i,\mathbf a_j)\bigr]\Bigr).
 \label{eq:pair-encoding}
\end{align*}
Both $f_n$ and $f_e$ apply a learned affine transformation, GELU~\cite{hendrycks2016gaussian}, and layer normalization~\cite{ba2016layer}. Figure~\ref{fig:syntax-encoding}A-B shows how these inputs enter the model.

\noindent\textbf{Message passing through shared symbols.}
We view symbols as 0-cells and candidate links as 1-cells whose boundaries are their two endpoints~\cite{hajij2022topological}. Candidates sharing a symbol exchange information through that symbol. For example, several possible notehead attachments to one slur contribute to the slur's context before their connection scores are updated (see Figure \ref{fig:syntax-encoding}C). Messages thus pass from candidates to symbols and back, between cells of different ranks, which we call hierarchical message passing.

For each candidate $e=\{i,j\}$, we write $(i,e)$ and $(j,e)$ for its incidences at the two endpoints. Each incidence acts as a candidate hyperedge that joins the candidate with both of its endpoints. For example, $(i,e)$ represents the candidate edge $e$ in the context of object $i$. The candidate hyperedge token $\boldsymbol{\tau}_{i,e}^{(t)}$ at message-passing step $t$ combines the current states of both symbols $\bigl(\mathbf{h}_i^{(t)},\mathbf{h}_j^{(t)}\bigr)$, the candidate edge state $\mathbf{z}_e^{(t)}$, and the incidence geometry $\boldsymbol\delta_{i\to j}$ from Table~\ref{tab:syntax-geometry},
\begin{equation*}
 \boldsymbol{\tau}_{i,e}^{(t)}=f_\tau\Bigl(\bigl[\mathbf h_i^{(t)}\Vert\mathbf h_j^{(t)}\Vert\mathbf z_e^{(t)}\Vert
 \boldsymbol\delta_{i\to j}\bigr]\Bigr),
\end{equation*}
where $f_\tau$ is a multilayer perceptron. The token $\boldsymbol{\tau}_{j,e}^{(t)}$ is computed in the same way with $i$ and $j$ exchanged. $K$ attention channels aggregate these tokens. For channel $k$, a learned vector $\mathbf w_k$ scores each token, and a softmax compares the candidates incident to symbol $i$,
\begin{equation*}
 \alpha_{i,e,k}^{(t)}=
 \frac{\exp(\mathbf w_k^\top\boldsymbol\tau_{i,e}^{(t)})}
 {\sum_{e'\in\mathcal I_i}\exp(\mathbf w_k^\top\boldsymbol{\tau}_{i,e'}^{(t)})}.
 \label{eq:incidence-attention}
\end{equation*}
Each channel takes a weighted sum of projected tokens. Concatenating the $K$ sums and applying a learned projection $f_c$ gives the symbol context,
\begin{equation*}
 \mathbf c_{i}^{(t)}=f_c\!\left(
 \mathop{\big\Vert}_{k=1}^{K}
 \sum_{e\in\mathcal I_i}\alpha_{i,e,k}^{(t)}V_k\boldsymbol{\tau}_{i,e}^{(t)}
 \right),
 \label{eq:target-context}
\end{equation*}
where $V_k$ is a learned affine projection.
The model updates every symbol in the same way from its own context, so $\mathbf h_j^{(t+1)}$ is computed similarly to $\mathbf h_i^{(t+1)}$. It then updates each candidate embedding $\mathbf{z}_e^{(t+1)}$ from its previous state, its two updated endpoints, their contexts, and the pair geometry $\mathbf r_e$,
\begin{align*}
 \mathbf h_i^{(t+1)}&=\operatorname{LN}\!\Bigl(\mathbf h_i^{(t)}+f_h\bigl([\mathbf h_i^{(t)}\Vert\mathbf c_i^{(t)}]\bigr)\Bigr),\\
 \mathbf u_e^{(t+1)}&=\Bigl[\mathbf z_e^{(t)}\Vert S(\mathbf h_i^{(t+1)},\mathbf h_j^{(t+1)})\Vert S(\mathbf c_{i}^{(t)},\mathbf c_{j}^{(t)})\Vert\mathbf r_e\Bigr],\\
 \mathbf z_e^{(t+1)}&=\operatorname{LN}\!\Bigl(\mathbf z_e^{(t)}+f_z\bigl(\mathbf u_e^{(t+1)}\bigr)\Bigr).
\end{align*}
Here $\operatorname{LN}$ is layer normalization and $f_h,f_z$ are multilayer perceptrons. Two rounds of these updates produce $\mathbf z_e^{(2)}$. The final logit combines the connection confidence logit $b_e$ with a topology correction,
\begin{align*}
 s_e&=b_e+f_{\rm base}(\mathbf z_e^{(0)})+f_{\rm topo}(\mathbf z_e^{(2)}),
 \qquad p_e=\sigma(s_e),
 \label{eq:edge-probability}
\end{align*}
where $\sigma$ is the sigmoid function, $f_{\rm base}$ and $f_{\rm topo}$ are two-layer perceptrons with scalar outputs, and $b_e+f_{\rm base}(\mathbf z_e^{(0)})$ is the base logit. Thus candidates compete when forming a shared context, while their final sigmoid probabilities allow several connections to the same symbol.

\noindent\textbf{Training objective.}
Let $y_e=1$ when the annotated syntax graph connects the two endpoints in either direction. The model is trained with binary cross-entropy $\mathcal L_{\rm edge}$ on the final logits and an auxiliary binary cross-entropy $\mathcal L_{\rm base}$ on the base logits. 
Both use all positive candidates and up to five negative candidates per positive, half chosen among those the model currently scores highest and the rest drawn at random.
We also supervise connection counts with $\mathcal L_{\rm count}$ and the ordering of competing candidates with $\mathcal L_{\rm rank}$,
\begin{equation*}
 \mathcal L=\mathcal L_{\rm edge}
 +\lambda_{\rm base}\mathcal L_{\rm base}
 +\lambda_{\rm count}\mathcal L_{\rm count}
 +\lambda_{\rm rank}\mathcal L_{\rm rank}.
\end{equation*}
where the hyperparameters $\lambda_{\rm base}$, $\lambda_{\rm count}$, and $\lambda_{\rm rank}$ weight the three auxiliary losses.

The count loss and the ranking loss group the candidates of each symbol by the class of the other endpoint. For example, all candidate links between a particular slur and noteheads of the same class form one group $\mathcal{B}$, and those that join the same slur to stems form another. For each group $\mathcal{B}$, let $m_\mathcal{B}=\sum_{e\in \mathcal{B}}y_e$ and $\widehat m_\mathcal{B}=\sum_{e\in \mathcal{B}}p_e$. The count penalty of $\mathcal{B}$ is
\begin{equation*}
 \ell_\mathcal{B}=\operatorname{SmoothL1}\!\left(
 \frac{\widehat m_\mathcal{B}-m_\mathcal{B}}{\sqrt{m_\mathcal{B}+1}}\right),
\end{equation*}
where $\operatorname{SmoothL1}$ is the smooth $L_1$ loss \cite{girshick2015fast}. The count loss takes the mean of $\ell_\mathcal{B}$ separately over groups with $m_\mathcal{B}>0$ and $m_\mathcal{B}=0$ and down-weights the latter by hyperparameter $\eta$,
\begin{equation*}
 \mathcal L_{\rm count}=\operatorname{mean}_{B:\,m_B>0}(\ell_B)
 +\eta\operatorname{mean}_{B:\,m_B=0}(\ell_B).
\end{equation*}
For each group containing both positive and negative candidates, the ranking loss encourages the lowest positive logit to exceed the highest negative logit by margin hyperparameter $\mu$,
\begin{align*}
 r_\mathcal{B}&=\operatorname{softplus}\Bigl(\mu-\min_{e\in \mathcal{B}:y_e=1}s_e+\max_{e\in \mathcal{B}:y_e=0}s_e\Bigr),\\
 \mathcal L_{\rm rank}&=\operatorname{mean}_{\mathcal{B}:\,0<m_\mathcal{B}<|\mathcal{B}|}(r_\mathcal{B}),
\end{align*}
where each mean runs over the groups that satisfy its condition and $\operatorname{softplus}(x)=\log(1+e^{x})$. Unlike the two cross-entropy terms, $\mathcal L_{\rm count}$ and $\mathcal L_{\rm rank}$ are computed over complete groups without sampling negatives. Table~\ref{tab:syntax-hparams} gives the values of the hyperparameters $\lambda_{\rm base}$, $\lambda_{\rm count}$, $\lambda_{\rm rank}$, $\eta$, and $\mu$, and Section~\ref{app:syntax-experiments} gives network dimensions, sampling settings, and optimization details.

\noindent\textbf{Direction and decoding.}
A logistic classifier trained on annotated directions chooses one orientation for each pair. Its inputs are the two symbol classes and signed box geometry, with endpoints ordered consistently by class and position. The model assigns $p_e$ to the chosen direction and zero to the reverse. Directed edges whose probability reaches a threshold chosen from $\{0.01,0.02,\ldots,0.99\}$ on the validation pages form the predicted syntax links.

\subsection{Time Groups and Precedence Links}
\label{sec:time-structure}
To convert a page of handwritten music notation to typeset notation, we need to know which notes in the reduction occur at the same time and the order of notes within each voice. This information allows us to determine the time position of each note and the structural levels of the voices (Section~\ref{sec:semantic-conversion}). In CWMN, both kinds of information can be derived from note durations. Widely used OMR datasets such as MUSCIMA++ annotate only syntax links~\cite{muscimadataset}, and learned notation-assembly methods also predict only syntax links~\cite{pacha2019learning,penarrubia2023efficient,yang2024complete}. In Schenkerian graphs, noteheads, stems, beams, and flags indicate structural levels rather than durations, so we treat time groups $\mathcal H$ and precedence links $\E_p$ as two separately predicted relations.

\noindent\textbf{Time groups.}
A time group is the set of noteheads that occur at the same time in the musical reduction. Let $(x_i,y_i)$ be the center of notehead $i$, and let $\bar w$ and $\bar h$ be the median notehead width and height on that page. We first estimate the probability that each pair of noteheads belongs to the same time group and then cluster the noteheads into groups based on these probabilities. Candidate pairs are noteheads $i$ and $j$ that are at most four notehead widths apart horizontally, $|x_i-x_j|\le 4\bar w$.

For each candidate pair, a histogram gradient-boosting classifier~\cite{pedregosa2011scikit} gives a probability $p_{ij}$, and noteheads that do not form a candidate pair have $p_{ij}=0$. The classifier takes three kinds of input.
First, its geometric features are the horizontal and vertical distances ($|x_i-x_j|/\bar w$ and $|y_i-y_j|/\bar h$, respectively) between the two noteheads, the same vertical distance ($|y_i-y_j|$) relative to the page height, the page aspect ratio, and the median notehead width and height relative to the page.
Second, its notational features are two binary indicators, one for whether the two noteheads touch the same stem and one for whether they are adjacent pitches drawn side by side, that is, $|x_i-x_j|\le 2\bar w$ and $|y_i-y_j|\le 1.5\bar h$. Third, its count features record the symbols that actually lie between the two noteheads. One counts the other noteheads that lie horizontally between them and no more than $3\bar h$ outside their vertical range. Another counts the noteheads stacked in the same column between them, namely those within $\bar w/2$ of their horizontal midpoint and vertically between them. Finally, a third counts the clefs whose vertical positions fall between the two noteheads. The classifier also receives the number of hollow noteheads in the pair and the total number of noteheads on the page.

Average-linkage clustering then produces the set of time groups $\mathcal H$. The average probability between two groups of noteheads $U$ and $U'$ is
\begin{equation*}
 \bar p(U,U')=\frac{1}{|U|\,|U'|}\sum_{i\in U}\sum_{j\in U'}p_{ij}.
\end{equation*}
Each notehead starts as its own group, and at every step the two groups with the largest $\bar p$ are merged. This continues until $\bar p$ between every two groups is below a threshold chosen on the validation pages.

\noindent\textbf{Precedence links.}
Precedence links connect each symbol to the one that follows it in its voice or hierarchical structure level, and in our annotations they always point forwards in time (to the right). The candidate successors of a symbol are therefore the symbols of the same family on its right, where noteheads, structural stems, and Arabic and Roman numerals each form one family. Here we again use a histogram gradient-boosting classifier to estimate the probability that each candidate directly follows the symbol. Its inputs include the relative position of the two symbols, the distance rank of each symbol among the other's candidates, the number of same-family symbols between them, the number of beams and slurs spanning both, and their stem directions. Each symbol keeps its highest-scoring candidate when that score exceeds a threshold. Where a voice divides, a second, higher threshold determines whether further successors are admitted. Both thresholds are chosen on the validation pages.

\subsection{Semantic Conversion}
\label{sec:semantic-conversion}
Algorithm~1 converts the semantic graph $\G$ into the JSON notation introduced by Ni-Hahn et al.~\cite{nihahn2024newdatasetnotationsoftware}, determining the time position, pitch, voice, and structural depth of each note.

\noindent\textbf{Time and pitch.}
Noteheads in the same time group share one time position, so we assign time positions by ordering the time groups according to the precedence links between their noteheads. Precedence links also connect structural stems, and we transfer the precedence between stems to their noteheads through the syntax links (line~1). The time groups are then topologically ordered by horizontal position, and the index of each group gives the time position of its noteheads (line~2). Pitch follows from the staff position of the notehead, the clef, the key signature, and attached accidentals (line~3).

\noindent\textbf{Voices.}
Noteheads form sequences along precedence links (line~4). The voice of a sequence depends on its first hollow notehead or beamed note. For a hollow notehead, a sequence with an upward stem belongs to the outer treble voice and one with a downward stem belongs to the outer bass voice. For a beamed note, a sequence that lies more often above the other sequences at shared times belongs to the outer treble voice, and one that lies more often below them belongs to the outer bass voice. Sequences with neither mark are inner voices, which we order by their average vertical position (lines~5--15).

\noindent\textbf{Depth.}
Notes without stems have depth zero. Stemmed notes are layered by the positions of their stem tips. In the upper voices we compare how high the tips reach, and in the lower voices how low they reach. A tip farther from its notehead marks a deeper level (lines~16--17). The two outer voices may receive different numbers of levels, for example, when one voice is much sparser than the other. We therefore align the treble and bass levels so that the deepest level agrees across all voices, which gives the hollow noteheads of the \textit{Ursatz} the same depth (line~18).

\begin{figure}[!t]
\small
\raggedright
\hrule
\vspace{4pt}
\textbf{Algorithm 1: Graph-to-JSON}\par
\renewcommand{\algorithmicrequire}{\textbf{Input:}}
\renewcommand{\algorithmicensure}{\textbf{Output:}}
\begin{algorithmic}[1]
\REQUIRE Page image and semantic graph $\G=(\V,\E_s,\E_p,\mathcal H)$, with noteheads $\mathcal N\subset\V$.
\ENSURE $\mathrm{time}[n]$, $\mathrm{pitch}[n]$, $\mathrm{voice}[n]$, and $\mathrm{depth}[n]$ for each notehead $n\in\mathcal N$.
\item[] \vspace{2pt}\textit{<<Time and Pitch>>}
\STATE For time ordering, map stem precedence to noteheads via $\E_s$.
\STATE Topologically order $\mathcal H$ based on horizontal position and assign group indices $\mathrm{time}[n]$.
\STATE Detect staff lines on the page image and infer $\mathrm{pitch}[n]$ from staff position, clef, key signature, and accidental.
\item[] \vspace{2pt}\textit{<<Voices from Structural Marks>>}
\STATE Trace sequences along the notehead precedence links in $\E_p$.
\FOR{each sequence $s$}
  \STATE $m\gets$ first hollow notehead or beamed note in $s$.
  \IF{$m$ is a hollow notehead}
    \STATE $\mathrm{role}[s]\gets$ outer treble/bass for an upward/downward stem, respectively.
  \ELSIF{$m$ is a beamed note}
    \STATE $\mathrm{role}[s]\gets$ outer treble/bass if $s$ lies more often above/below other sequences at shared times.
  \ELSE
    \STATE $\mathrm{role}[s]\gets$ inner voice.
  \ENDIF
\ENDFOR
\STATE Order inner voices by average relative vertical position, and set $\mathrm{voice}[n]$ to $\mathrm{role}[s]$ of the sequence $s$ containing $n$.
\item[] \vspace{2pt}\textit{<<Depth from Stem Tips>>}
\STATE $\mathrm{depth}[n]\gets0$ for stemless notes.
\STATE Assign levels by relative stem-tip height above upper and below lower voices.
\STATE Align treble and bass levels.
\STATE \textbf{return} $(\mathrm{time},\mathrm{pitch},\mathrm{voice},\mathrm{depth})$.
\end{algorithmic}
\vspace{4pt}
\hrule
\vspace{-2.5mm}
\end{figure}

\section{Experiments}
\label{sec:experiments}

\subsection{Symbol Recognition}
\label{sec:symbol-experiments}
\begin{table*}[!t]
\centering
\caption{Symbol detection on handwritten Schenkerian analysis. All values are percentages. The upper block gives the mean $\pm$ sample standard deviation (SD) across the eight random splits, and the lower block gives the mean $\pm$ sample SD over two runs on the test pages of the held-out work. Bold indicates the highest value in each column of each block.}
\label{tab:symbol-main-results}
\footnotesize
\setlength{\tabcolsep}{1.6pt}
\begin{tabular*}{\textwidth}{@{\extracolsep{\fill}}lcccccccc@{}}
\toprule
 & \multicolumn{4}{c}{All symbols} & \multicolumn{2}{c}{\shortstack{Large symbols\\$L>400$ pixels}} & \multicolumn{2}{c}{\shortstack{Very large symbols\\$L>800$ pixels}} \\
\cmidrule(lr){2-5}\cmidrule(lr){6-7}\cmidrule(l){8-9}
Method & mAP$_{50}$ & w-mAP$_{50}$ & mAP$_{50:95}$ & w-mAP$_{50:95}$ & mAP$_{50}$ & w-mAP$_{50}$ & mAP$_{50}$ & w-mAP$_{50}$ \\
\midrule
\multicolumn{9}{c}{\textbf{Eight random splits}} \\
\midrule
\multicolumn{9}{c}{\textit{Baselines}} \\
Faster R-CNN + Inception-ResNet-v2~\cite{pacha2018handwritten} & 58.37$\pm$3.12 & 87.13$\pm$2.21 & 33.94$\pm$2.06 & 47.92$\pm$2.60 & 30.84$\pm$2.80 & 55.96$\pm$3.51 & 31.87$\pm$2.82 & 44.05$\pm$3.35 \\
completeOMR~\cite{yang2024complete} & 55.25$\pm$2.21 & 87.77$\pm$1.03 & 40.41$\pm$1.58 & 60.62$\pm$0.86 & 44.00$\pm$3.84 & 69.97$\pm$2.61 & 40.30$\pm$4.65 & 54.52$\pm$4.84 \\
MuSViT + Faster R-CNN~\cite{penarrubia2026musvit} & 62.81$\pm$2.89 & 91.03$\pm$1.31 & 37.93$\pm$1.84 & 54.02$\pm$0.89 & 49.20$\pm$2.54 & 71.13$\pm$2.20 & 45.28$\pm$5.64 & 52.70$\pm$4.68 \\
\addlinespace[3pt]
\multicolumn{9}{c}{\textit{Ours}} \\
YOLOv8L only & 62.46$\pm$2.80 & 92.99$\pm$0.86 & 46.89$\pm$2.08 & 67.28$\pm$1.53 & 43.37$\pm$4.30 & 70.17$\pm$1.63 & 43.18$\pm$4.27 & 54.56$\pm$3.72 \\
RF-DETR only & 48.40$\pm$2.42 & 86.66$\pm$1.46 & 32.28$\pm$1.59 & 54.90$\pm$1.11 & \textbf{58.29}$\pm$4.30 & 83.17$\pm$2.31 & \textbf{67.75}$\pm$5.44 & \textbf{81.05}$\pm$4.07 \\
\textbf{YOLOv8L + RF-DETR} & \textbf{62.89}$\pm$2.82 & \textbf{93.47}$\pm$0.81 & \textbf{47.53}$\pm$2.09 & \textbf{67.81}$\pm$1.50 & 54.50$\pm$3.11 & \textbf{84.56}$\pm$1.65 & 59.07$\pm$6.12 & 80.58$\pm$3.97 \\
\midrule
\multicolumn{9}{c}{\textbf{Held-out work}} \\
\midrule
\multicolumn{9}{c}{\textit{Baselines}} \\
Faster R-CNN + Inception-ResNet-v2~\cite{pacha2018handwritten} & 68.03$\pm$3.47 & 92.51$\pm$0.46 & 41.96$\pm$1.98 & 61.80$\pm$1.38 & 27.70$\pm$1.50 & 52.21$\pm$3.81 & 0.00$\pm$0.00 & 0.00$\pm$0.00 \\
completeOMR~\cite{yang2024complete} & 63.40$\pm$1.05 & 93.86$\pm$0.07 & 46.11$\pm$0.42 & 69.45$\pm$0.67 & 39.09$\pm$3.77 & 71.03$\pm$1.19 & 17.45$\pm$0.46 & 30.97$\pm$0.21 \\
MuSViT + Faster R-CNN~\cite{penarrubia2026musvit} & 71.19$\pm$1.88 & 95.35$\pm$0.01 & 44.85$\pm$1.26 & 63.13$\pm$0.01 & 40.63$\pm$0.30 & 67.74$\pm$1.96 & 21.59$\pm$7.00 & 36.59$\pm$8.80 \\
\addlinespace[3pt]
\multicolumn{9}{c}{\textit{Ours}} \\
YOLOv8L only & 71.95$\pm$3.45 & 95.83$\pm$0.20 & 53.31$\pm$2.77 & 73.11$\pm$0.27 & 33.02$\pm$3.84 & 62.96$\pm$0.98 & 6.12$\pm$2.46 & 12.82$\pm$2.89 \\
RF-DETR only & 57.91$\pm$0.01 & 88.92$\pm$0.29 & 39.62$\pm$1.56 & 57.62$\pm$0.80 & 58.83$\pm$0.28 & 87.54$\pm$0.12 & \textbf{71.06}$\pm$5.56 & \textbf{84.48}$\pm$1.91 \\
\textbf{YOLOv8L + RF-DETR} & \textbf{73.53}$\pm$3.15 & \textbf{96.74}$\pm$0.23 & \textbf{54.87}$\pm$2.38 & \textbf{74.09}$\pm$0.32 & \textbf{60.97}$\pm$2.30 & \textbf{87.89}$\pm$1.43 & 70.25$\pm$5.01 & 79.55$\pm$0.25 \\
\bottomrule
\end{tabular*}
\par\smallskip
\begin{minipage}{\textwidth}
\footnotesize
$L$ is the longest side of a symbol's bounding box in pixels after the page is scaled to $5100\times3509$.
\end{minipage}
\vspace{-5mm}
\end{table*}

\begin{table*}[!t]
\centering
\caption{Symbol detection on MUSCIMA++. All methods use the 73-class vocabulary and the 84/28/28-page split of completeOMR~\cite{yang2024complete}. All values are percentages, given as the mean $\pm$ sample SD over two runs. Bold indicates the highest value in each column.}
\label{tab:muscima-symbol-size}
\footnotesize
\setlength{\tabcolsep}{1.8pt}
\begin{tabular*}{\textwidth}{@{\extracolsep{\fill}}lcccccccc@{}}
\toprule
 & \multicolumn{4}{c}{All symbols} & \multicolumn{2}{c}{\shortstack{Large symbols\\$L>400$ pixels}} & \multicolumn{2}{c}{\shortstack{Very large symbols\\$L>800$ pixels}} \\
\cmidrule(lr){2-5}\cmidrule(lr){6-7}\cmidrule(l){8-9}
Method & mAP$_{50}$ & w-mAP$_{50}$ & mAP$_{50:95}$ & w-mAP$_{50:95}$ & mAP$_{50}$ & w-mAP$_{50}$ & mAP$_{50}$ & w-mAP$_{50}$ \\
\midrule
\multicolumn{9}{c}{\textit{Baselines}} \\
\midrule
Faster R-CNN + Inception-ResNet-v2~\cite{pacha2018handwritten} & 70.94$\pm$1.82 & 90.65$\pm$0.13 & 46.88$\pm$0.86 & 58.03$\pm$0.46 & 40.14$\pm$9.64 & 34.19$\pm$3.13 & 15.86$\pm$1.07 & 5.81$\pm$0.82 \\
completeOMR~\cite{yang2024complete} & 84.24$\pm$0.55 & 96.73$\pm$0.09 & 69.86$\pm$0.28 & 72.38$\pm$0.92 & 79.19$\pm$2.16 & 87.47$\pm$0.16 & \textbf{62.41}$\pm$1.49 & 72.00$\pm$1.32 \\
MuSViT + Faster R-CNN~\cite{penarrubia2026musvit} & 84.53$\pm$0.72 & 95.28$\pm$0.46 & 59.67$\pm$1.57 & 67.05$\pm$0.34 & 79.62$\pm$7.19 & 80.35$\pm$2.80 & 60.36$\pm$0.35 & 61.87$\pm$0.55 \\
\midrule
\multicolumn{9}{c}{\textit{Ours}} \\
\midrule
YOLOv8L only & \textbf{87.44}$\pm$0.86 & 97.81$\pm$0.04 & \textbf{80.02}$\pm$0.56 & 86.04$\pm$0.21 & 78.88$\pm$2.16 & 85.42$\pm$0.66 & 53.72$\pm$1.17 & 65.91$\pm$1.99 \\
RF-DETR only & 60.57$\pm$0.69 & 91.74$\pm$0.03 & 46.38$\pm$0.66 & 68.65$\pm$0.12 & 66.72$\pm$0.40 & 87.10$\pm$0.16 & 56.94$\pm$3.79 & 79.27$\pm$0.87 \\
\textbf{YOLOv8L + RF-DETR} & 87.13$\pm$1.08 & \textbf{97.93}$\pm$0.02 & 79.69$\pm$0.68 & \textbf{86.06}$\pm$0.28 & \textbf{82.48}$\pm$0.71 & \textbf{92.49}$\pm$0.95 & 60.65$\pm$1.70 & \textbf{85.50}$\pm$0.36 \\
\bottomrule
\end{tabular*}
\par\smallskip
\begin{minipage}{\textwidth}
\footnotesize
$L$ is the longest side of a symbol's bounding box in pixels after the page is scaled to $5100\times3509$.
\end{minipage}
\vspace{-5mm}
\end{table*}

We evaluate symbol recognition on our handwritten Schenkerian dataset and on MUSCIMA++~\cite{muscimadataset}, which contains conventional handwritten scores. Section~\ref{sec:dataset-splits} describes our dataset splits, and Section~\ref{app:symbol-hparams} gives the training settings. Unlike the detector used for data collection in Section~\ref{sec:data-collection}, the detectors compared here are not pretrained on MUSCIMA++. They start from weights pretrained on COCO~\cite{lin2014microsoft} and are trained directly on each dataset, while MuSViT + Faster R-CNN starts from the public MuSViT encoder~\cite{penarrubia2026musvit}, which is pretrained on sheet music that includes 9.7 million International Music Score Library Project (IMSLP) pages.

We first match predicted boxes to ground-truth boxes and then compute the
metrics. At each intersection over union (IoU) threshold, the one-to-one
matcher of Ultralytics~\cite{yolov8_ultralytics} pairs boxes of the same
class, favoring larger IoU values. This gives the average precision (AP) of
each class, the area under the precision-recall curve, integrated over 101
recall points. AP$_{50}$ uses a threshold of 0.5, and AP$_{50:95}$ averages
the ten COCO~\cite{lin2014microsoft} thresholds from 0.5 to 0.95 in steps of
0.05. We take mAP as the equal mean of AP over all classes with ground-truth
instances, while w-mAP weights each class by its number of instances.

As in Section~\ref{sec:symbol-recognition}, we measure symbol size by $L$ after scaling every page of both datasets to $5100\times3509$, and report two large-symbol subsets separately, $L>400$
and $L>800$. Matching in these subsets follows COCO~\cite{lin2014microsoft}. Predictions are
processed in descending confidence order, and each goes to the unused
ground-truth box of the same class with the greatest IoU of at least 0.5,
with boxes of the required size taking priority. Predictions matched to
boxes outside the size range are ignored. An unmatched prediction counts as
a false positive only if its own size exceeds the cutoff. All detectors use the same rules.

Tables~\ref{tab:symbol-main-results} and~\ref{tab:muscima-symbol-size} show
that although RF-DETR comes close to or even surpasses our full model,
YOLOv8L + RF-DETR, on the large-symbol metrics of some datasets, our full
model performs best overall and is the strongest on both w-mAP$_{50}$ and
w-mAP$_{50:95}$.
Table~\ref{tab:muscima-symbol-size} shows that our approach does well not
only on our own handwritten Schenkerian dataset but also on the external
MUSCIMA++ dataset, where it leads on five of the eight metrics, including
three of the four large-symbol metrics. YOLOv8L alone is slightly higher on
mAP$_{50}$ and mAP$_{50:95}$ over all symbols, while completeOMR is higher
on mAP$_{50}$ for $L>800$. MUSCIMA++ contains very tall
symbols, such as barlines that span all the staves of a system and exceed the
height of our strips, so the RF-DETR branch takes whole pages on this
dataset.

\subsection{Syntax-Link Prediction}
\label{sec:syntax-experiments}
We evaluate the syntax-link model of Section~\ref{sec:syntax-predictor} on MUSCIMA++~\cite{muscimadataset}, DoReMi~\cite{shatri2021doremi}, and \ourdataset{}. We compare it with the MLP32 and MLP64--512 models from Efficient Notation Assembly (Efficient NA)~\cite{penarrubia2023efficient}, the MLP from completeOMR~\cite{yang2024complete}, logistic regression~\cite{hastie2009elements}, histogram gradient boosting (HGB)~\cite{pedregosa2011scikit}, a pair MLP, and a two-layer GraphSAGE-based model~\cite{hamilton2017graphsage}. Section~\ref{app:syntax-hparams} gives the training settings. We report directed-edge micro-F1, which adds up true positives, false positives, and false negatives over all test pages before computing a single F1. Ground-truth edges outside the candidate set count as false negatives.

MUSCIMA++ and DoReMi use splits that hold out entire works. MUSCIMA++ contains 84/28/28 pages from 12/4/4 works, and we also report results excluding links with an endpoint of a staff class, namely staff, staff line, staff space, or staff group. DoReMi uses 423/140/216 pages of musical works. \ourdataset{} uses the eight random splits and the held-out split of Section~\ref{sec:dataset-splits}.

\begin{table}[!t]
\centering
\caption{Directed syntax-edge micro-F1 (\%, mean $\pm$ sample SD over two runs) on MUSCIMA++ and DoReMi. The non-staff column excludes staff-related links.}
\label{tab:syntax-edges}
\footnotesize
\setlength{\tabcolsep}{1.5pt}
\begin{tabular*}{\linewidth}{@{\extracolsep{\fill}}lccc@{}}
\toprule
 & \multicolumn{2}{c}{MUSCIMA++} & DoReMi \\
\cmidrule(lr){2-3}\cmidrule(l){4-4}
Method & All links & Non-staff links & All links \\
\midrule
\multicolumn{4}{@{}l}{\textit{Baselines}} \\
Efficient NA--MLP32~\cite{penarrubia2023efficient} & $87.66{\pm}5.35$ & $82.81{\pm}5.71$ & $85.21{\pm}0.73$ \\
Efficient NA--MLP64--512~\cite{penarrubia2023efficient} & $86.88{\pm}7.53$ & $78.36{\pm}13.64$ & $83.67{\pm}0.94$ \\
completeOMR--MLP~\cite{yang2024complete} & $90.56{\pm}0.79$ & $86.19{\pm}0.58$ & $81.11{\pm}0.12$ \\
\multicolumn{4}{@{}l}{\textit{General-purpose models}} \\
Logistic regression & $70.30{\pm}0.19$ & $60.65{\pm}0.24$ & $71.10{\pm}0.29$ \\
HGB & $94.39{\pm}0.07$ & $90.95{\pm}0.08$ & $88.44{\pm}0.08$ \\
Pair MLP & $94.19{\pm}0.04$ & $90.66{\pm}0.02$ & $89.18{\pm}0.39$ \\
GraphSAGE & $93.96{\pm}0.15$ & $90.36{\pm}0.20$ & $89.45{\pm}0.29$ \\
\midrule
\textbf{Ours} & $\mathbf{96.68{\pm}0.46}$ & $\mathbf{94.69{\pm}0.50}$ & $\mathbf{94.11{\pm}0.48}$ \\
\bottomrule
\end{tabular*}
\vspace{-3mm}
\end{table}

\begin{table}[!t]
\centering
\caption{Syntax-link prediction on \ourdataset{}. Directed-edge micro-F1 (\%), with the sample SD across the eight random splits and across two runs on the held-out work.}
\label{tab:schenker-syntax-edges}
\footnotesize
\setlength{\tabcolsep}{3pt}
\begin{tabular*}{\linewidth}{@{\extracolsep{\fill}}lcc@{}}
\toprule
Method & \shortstack{Mean of eight\\random splits} & \shortstack{Held-out\\work} \\
\midrule
\multicolumn{3}{@{}l}{\textit{Baselines}} \\
Efficient NA--MLP32~\cite{penarrubia2023efficient} & $51.10{\pm}3.24$ & $44.67{\pm}1.33$ \\
Efficient NA--MLP64--512~\cite{penarrubia2023efficient} & $64.54{\pm}3.04$ & $60.47{\pm}3.23$ \\
completeOMR--MLP~\cite{yang2024complete} & $51.64{\pm}2.13$ & $44.76{\pm}3.33$ \\
\multicolumn{3}{@{}l}{\textit{General-purpose models}} \\
Logistic regression & $39.02{\pm}2.96$ & $33.71{\pm}0.12$ \\
HGB & $76.61{\pm}1.76$ & $67.80{\pm}1.57$ \\
Pair MLP & $66.29{\pm}4.05$ & $59.95{\pm}0.87$ \\
GraphSAGE & $75.49{\pm}2.34$ & $69.64{\pm}2.87$ \\
\midrule
\textbf{Ours} & $\mathbf{86.48{\pm}1.82}$ & $\mathbf{85.70{\pm}0.37}$ \\
\bottomrule
\end{tabular*}
\vspace{-3.5mm}
\end{table}

Table~\ref{tab:syntax-edges} shows that our model achieves the highest F1 among the compared methods. Its lead on MUSCIMA++ grows when staff-related links are excluded, so the improvement extends to relations between musical symbols beyond their attachment to the staff. The advantage is largest on Schenkerian analysis, where Table~\ref{tab:schenker-syntax-edges} shows that our model leads clearly on the random splits and on the held-out work. Combining the classes and boxes of each candidate pair with context exchanged between symbols and candidate links is effective on both conventional notation and Schenkerian analysis.

\begin{figure*}[!t]
\centering
\includegraphics[width=\textwidth]{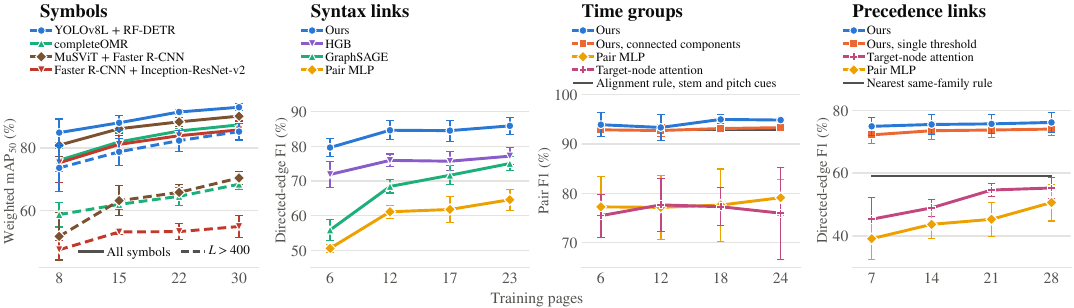}
\caption{Effect of training-data size on the four tasks. The horizontal axis gives the number of training pages, which correspond to 25\%, 50\%, 75\%, and 100\% of the training pages. Points are means over three of the random splits and bars are sample SDs. In the symbol panel, solid lines show all symbols and dashed lines show $L>400$. Horizontal lines mark rules, whose results do not change with the number of training pages.}
\label{fig:data-size}
\vspace{-2.5mm}
\end{figure*}

\subsection{Time Groups and Precedence Links}
\label{sec:time-experiments}
We score time groups by pair F1. Every two noteheads in the same group form an unordered pair, and we compare the predicted pairs with the ground-truth pairs, pooling counts over all test pages. Precedence links are scored by the same directed-edge micro-F1 as syntax links (Section~\ref{sec:syntax-experiments}).

Simple rules serve as baselines for both tasks. For time groups, the first rule puts noteheads in one group when their centers fall within a horizontal tolerance in notehead widths and a vertical tolerance relative to the page height. The second rule measures the vertical tolerance in notehead heights and also joins noteheads on a shared stem as well as adjacent pitches drawn side by side. For precedence, the rule links each symbol to the nearest symbol of the same family on its right. We also train a pair MLP and a target-node attention model that score every pair of symbols from symbol classes and box geometry. The latter also lets the candidate pairs that point to the same symbol attend to one another. Their time groups are the connected components of pairs above a threshold.

\begin{table}[!t]
\centering
\caption{Time-group prediction on \ourdataset{}. Pair F1 (\%), with the sample SD across the eight random splits and across two runs on the held-out work.}
\label{tab:time-groups}
\footnotesize
\setlength{\tabcolsep}{3pt}
\begin{tabular*}{\linewidth}{@{\extracolsep{\fill}}lcc@{}}
\toprule
Method & \shortstack{Mean of eight\\random splits} & \shortstack{Held-out\\work} \\
\midrule
Alignment rule, page-height tolerance & $87.79{\pm}6.55$ & 94.57 \\
Alignment rule, stem and pitch cues & $92.92{\pm}1.61$ & 93.73 \\
Pair MLP & $78.61{\pm}4.16$ & $92.10{\pm}1.25$ \\
Target-node attention & $77.03{\pm}7.89$ & $94.26{\pm}1.18$ \\
Ours, connected components & $93.44{\pm}1.48$ & $90.80{\pm}0.00^{\dagger}$ \\
\midrule
\textbf{Ours} & $\mathbf{94.98{\pm}1.33}$ & $\mathbf{97.54{\pm}0.00}^{\dagger}$ \\
\bottomrule
\end{tabular*}
\par\vspace{4pt}
\begin{minipage}{\linewidth}
\footnotesize
$^{\dagger}$The two runs on the held-out work use random seeds 42 and 43 and give identical results.
\end{minipage}
\end{table}

\begin{table}[!t]
\centering
\caption{Precedence-link prediction on \ourdataset{}. Directed-edge micro-F1 (\%), with the sample SD across the eight random splits and across two runs on the held-out work.}
\label{tab:precedence}
\footnotesize
\setlength{\tabcolsep}{3pt}
\begin{tabular*}{\linewidth}{@{\extracolsep{\fill}}lcc@{}}
\toprule
Method & \shortstack{Mean of eight\\random splits} & \shortstack{Held-out\\work} \\
\midrule
Nearest same-family rule & $57.01{\pm}3.35$ & 50.04 \\
Pair MLP & $50.32{\pm}4.15$ & $53.80{\pm}4.42$ \\
Target-node attention & $54.32{\pm}3.51$ & $56.85{\pm}3.53$ \\
Ours, single threshold & $72.20{\pm}2.06$ & $71.46{\pm}0.51$ \\
\midrule
\textbf{Ours} & $\mathbf{74.70{\pm}2.25}$ & $\mathbf{73.84{\pm}0.48}$ \\
\bottomrule
\end{tabular*}
\vspace{-5mm}
\end{table}

Table~\ref{tab:time-groups} shows that our time-group model gives the best pair F1 on the random splits and on the held-out work. For precedence links, Table~\ref{tab:precedence} shows that our model clearly outperforms the rule and both neural baselines. The decoding step matters in both tasks. Average linkage gives better time groups than connected components, and selecting successors per symbol gives better precedence links than keeping every pair above a single threshold.

\subsection{End-to-End Evaluation}
\label{sec:end-to-end}
\begin{table}[!t]
\centering
\caption{End-to-end notation assembly on the held-out work. F1 (\%) with annotated and with detected symbols, mean $\pm$ sample SD over two runs.}
\label{tab:end-to-end}
\footnotesize
\setlength{\tabcolsep}{3pt}
\begin{tabular*}{\linewidth}{@{\extracolsep{\fill}}lcc@{}}
\toprule
Task & \shortstack{Annotated\\symbols} & \shortstack{Detected\\symbols} \\
\midrule
Syntax links (directed-edge F1) & $85.70{\pm}0.37$ & $80.79{\pm}0.41$ \\
Time groups (pair F1) & $97.54{\pm}0.00$ & $95.04{\pm}0.18$ \\
Precedence links (directed-edge F1) & $73.84{\pm}0.48$ & $71.38{\pm}0.64$ \\
\bottomrule
\end{tabular*}
\end{table}

The experiments above give the relation models annotated symbols. To measure how errors in symbol recognition carry over to the later stages, we also apply the three models to the symbols that our detector finds on the test pages of the held-out work, keeping the models and their validation-selected settings unchanged. A predicted relation counts as correct only if both of its endpoints match annotated symbols of the same class with an IoU of at least 0.5 and the two matched symbols are joined by an annotated relation. Relations of missed symbols therefore count as false negatives.

Table~\ref{tab:end-to-end} shows that the full pipeline retains most of its accuracy when it starts from detected symbols. Even with detected symbols, each of our models still outperforms every baseline on the held-out work in Tables~\ref{tab:schenker-syntax-edges}, \ref{tab:time-groups}, and \ref{tab:precedence}, although those baselines receive annotated symbols.

\subsection{Scalability}
\label{sec:scalability}
To study how our method scales with the amount of training data, we retrain the models on 25\%, 50\%, and 75\% of the training pages of three of the random splits and keep the validation and test sets fixed. The 100\% points use the full training sets of the same three splits, and Figure~\ref{fig:data-size} shows all four tasks.

Symbol recognition depends most on the amount of data. The w-mAP$_{50}$ score rises steadily with more pages, and our method is the highest at every size. The advantage is larger for large symbols. With half of the pages, our method already exceeds completeOMR, MuSViT + Faster R-CNN, and Faster R-CNN + Inception-ResNet-v2 trained on all pages in w-mAP$_{50}$ for $L>400$. Our method is also the best at every size on syntax links, time groups, and precedence links. Syntax links also gain from more pages, while time groups and precedence links reach nearly the same F1 with a quarter of the pages. Symbol recognition is still rising at the full training set and can keep improving as more data becomes available.

\section{Training Settings}
\label{app:syntax-experiments}

\subsection{Symbol Recognition}
\label{app:symbol-hparams}

\begin{table}[!t]
\centering
\caption{Training settings of our symbol detectors. Where they differ, settings are listed separately for \ourdataset{} and MUSCIMA++.}
\label{tab:symbol-hparams}
\scriptsize
\setlength{\tabcolsep}{3pt}
\begin{tabular}{@{}>{\raggedright\arraybackslash}p{0.21\columnwidth}>{\raggedright\arraybackslash}p{0.33\columnwidth}>{\raggedright\arraybackslash}p{0.41\columnwidth}@{}}
\toprule
Setting & YOLOv8L branch (ours) & RF-DETR branch (ours) \\
\midrule
Training crops & $1216\times1216$ tiles, stride 960 & Full-width strips of height 768, stride 384 (\ourdataset{}), whole pages (MUSCIMA++) \\
Training targets & Fully contained targets & $L>200$, visible area at least 90\% \\
Optimizer & AdamW & AdamW \\
Learning rate & $3\times10^{-4}$ & $10^{-4}$ (detector), $1.5\times10^{-4}$ (encoder) \\
Batch size & 16  & 9  \\
Training length & At most 120 epochs & At most 120 epochs \\
Patience & 15 (\ourdataset{}), 20 (MUSCIMA++) epochs & 20 epochs \\
Initialization & COCO & COCO \\
Random seeds & 20261021 and 20261104 (\ourdataset{}), 20261027 and 20261104 (MUSCIMA++) & 42 and 43 (\ourdataset{}), 20261027 and 20261104 (MUSCIMA++) \\
\bottomrule
\end{tabular}
\vspace{-5mm}
\end{table}

Table~\ref{tab:symbol-hparams} lists the training settings of symbol detectors, and all hyperparameters are selected on validation data. completeOMR merges its predictions with the authors' code. We implement MuSViT + Faster R-CNN following its paper and public training script~\cite{penarrubia2026musvit}, placing a feature pyramid and Faster R-CNN on the MuSViT encoder adapted with rank-8 low-rank adaptation (LoRA)~\cite{hu2022lora}. It predicts on the tiles of our YOLOv8L and merges them in the same way, and MUSCIMA++ pages are inverted to dark ink on a light background for its encoder.

\subsection{Syntax-Link Prediction}
\label{app:syntax-hparams}

Table~\ref{tab:syntax-hparams} gives the settings of our model and our baselines.

\begin{table}[!t]
\centering
\caption{Hyperparameters of our syntax-link model and the pair MLP and GraphSAGE baselines. All models use random seeds 42 and 43.}
\label{tab:syntax-hparams}
\scriptsize
\setlength{\tabcolsep}{3pt}
\begin{tabular}{@{}>{\raggedright\arraybackslash}p{0.29\columnwidth}>{\raggedright\arraybackslash}p{0.47\columnwidth}>{\raggedright\arraybackslash}p{0.18\columnwidth}@{}}
\toprule
 & Ours & Pair MLP / GraphSAGE \\
\midrule
Class embedding & 32 & 24 \\
Hidden width & 128 & 96 \\
Layers & 2 incidence-attention, $K=4$ channels & 2 GraphSAGE \\
Dropout & 0.1 & 0.1 \\
Crop height $\times$ width, margin & $128\times256$, 15\% of the union box (at least 8 pixels) & -- \\
Optimizer, learning rate & AdamW, $8\times10^{-4}$ & AdamW, $10^{-3}$ \\
Learning-rate schedule & cosine decay & constant \\
Weight decay & $10^{-4}$ & $10^{-4}$ \\
Gradient-norm clipping & 2 & 5 \\
Label smoothing & 0.05 & -- \\
Maximum epochs & 80 & 80 \\
Patience & 12 & 12 \\
Loss weights & $\lambda_{\rm base}=0.25$, $\lambda_{\rm count}=0.15$, $\lambda_{\rm rank}=0.05$, $\eta=0.25$, $\mu=0.2$ & -- \\
\bottomrule
\end{tabular}
\end{table}

Each crop of our model covers the union of the two endpoint boxes. Every CNN block applies a $3\times3$ convolution, batch normalization, ReLU, and $2\times2$ max pooling, and the five blocks have 16, 32, 64, 128, and 128 channels. Following Pacha et al.~\cite{pacha2019learning}, we first train the CNN as a pairwise link classifier whose crops also contain the page image as a third channel, which is left empty when the frozen CNN encodes candidates for our model. Each dataset and each split gets its own CNN, trained from random initialization only on its training pages for 40 epochs, using Adam with learning rate $10^{-3}$, cosine decay, batch size 256, and label smoothing 0.1. Each epoch samples up to 25,000 positive pairs and five times as many negatives among pairs whose box gap does not exceed the largest gap of an annotated link between the same two classes. For the edge and base losses of the incidence model, each positive comes with up to five sampled negatives, half with the highest current scores and the rest drawn uniformly. The two losses draw separately, and pages without positive candidates use all negatives. The direction classifier receives one-hot endpoint classes and 15 standardized geometric features, namely center coordinates, signed displacements, box dimensions, log size ratios, the box gap relative to the page diagonal, and overlap ratios. It is fitted on training links with one annotated direction, and a probability of at least 0.5 points the link from the first endpoint to the second, with endpoints ordered by class and position.

We reimplement MLP32 and MLP64--512 from Efficient Notation Assembly~\cite{penarrubia2023efficient} and the MLP of completeOMR~\cite{yang2024complete}, which receives ground-truth symbols here, with the settings of their papers and train them on directed labels of our splits. MLP32 has one hidden layer of 32 units, and MLP64--512 has two hidden layers of 64 and 512 units. Logistic regression and HGB use the \texttt{SGDClassifier} and \texttt{HistGradient\allowbreak Boosting\allowbreak Classifier} of scikit-learn~\cite{pedregosa2011scikit}. The pair MLP builds each symbol representation from a class embedding and eight box features, then concatenates the two endpoint representations with 32 pair features and scores the result. These pair features are the eight box features of each endpoint and 16 directed offset, gap, overlap, and size-ratio features. GraphSAGE adds neighbor mean aggregation with residual connections and layer normalization before this pair encoder. The general-purpose models are trained with binary cross-entropy on every positive candidate and up to eight uniformly sampled negatives per positive on each page, and at least 512 when the page has few positives. The neural models redraw these negatives each epoch, while logistic regression and HGB keep one fixed sample.

We select checkpoints by validation directed-edge average precision and choose the threshold that maximizes pooled validation micro-F1, breaking ties by recall and then by the larger threshold. All predictors follow these criteria. The MUSCIMA++ non-staff evaluation drops links with a staff-related endpoint and keeps the all-link threshold.

\subsection{Time Groups and Precedence Links}
\label{app:temporal-hparams}

Both classifiers of Section~\ref{sec:time-structure} use the \texttt{HistGradient\allowbreak Boosting\allowbreak Classifier} of scikit-learn~\cite{pedregosa2011scikit} with learning rate 0.05, at most 31 leaf nodes, L2 regularization 1.0, and no early stopping, and they run for 300 iterations on time groups and 400 on precedence links. The tolerances of the rule baselines are also chosen on the validation pages. The pair MLP and target-node attention baselines use the encoder and training settings of the pair MLP in Table~\ref{tab:syntax-hparams}, and target-node attention further refines each pair representation by attending over the other candidates that share its target.

\FloatBarrier
\section{Conclusion}
\label{sec:conclusion}

In this paper, we introduce several novel contributions improving OMR modeling and the collection of semantic music data. We built \ourdataset{}, the first labeled dataset of handwritten Schenkerian analyses, containing 47 pages of analyses. The data were collected through a model-assisted, multi-stage process that repeatedly updated the models with completed pages and gave initial predictions for the next batch of pages. Building on this dataset, we proposed a recognition-to-typesetting framework involving an ensemble model for symbol detection and classification, a topological neural network for syntax links, and gradient-boosting classifiers for time groups and precedence links. Furthermore, we introduce an algorithm to convert semantic graphs into typeset Schenkerian notation. Our experiments show that every learned stage of our framework outperforms baseline methods in most metrics. Additionally, our experiments show that symbol recognition was still rising when all training pages were used. We therefore plan to extend \ourdataset{} with more analysts, more works, and printed and mixed-notation Schenkerian graphs so that Schenkerian OMR systems can provide a greater breadth of computer-readable music semantic data.

\section*{Acknowledgment of AI Assistance}
We used Claude Opus 5.5 and GPT-6 astra as AI assistants in this work. The research questions and ideas of this paper were proposed by the authors, and some of the ideas emerged from discussions with these assistants. The dataset and its annotation scheme were created by the authors. The experiments were designed by the authors, and the AI assistants helped write the code, schedule the training jobs, and compute the experimental results. All results were manually reviewed and checked by the authors. The search for and organization of related work were done with the help of the AI assistants. For the vast majority of the main text, the authors wrote the first draft, which was polished by Claude Opus 5.5 and then further revised by the authors. Some passages were written in Chinese, translated into English by Claude Opus 5.5, and then revised. Figure~\ref{fig:data-size} and all result tables were generated by the AI assistants and reviewed by the authors, and all other figures were drawn entirely by the authors. Bibliography entries were added with the help of the AI assistants and checked by the authors. We take full responsibility for the final content of this work, including text, claims, and artifacts produced with the aid of generative AI.

\FloatBarrier
\bibliographystyle{IEEEtran}
\bibliography{icde2027/references}
\end{document}